\documentclass[preprint,showpacs,preprintnumbers,amsmath,amssymb,12pt]{revtex4}
\usepackage{graphicx}% Include figure files
\usepackage{dcolumn}% Align table columns on decimal point
\usepackage{bm}% bold math
\usepackage{amssymb}
\usepackage{amsmath}
\usepackage{latexsym}

\begin{document}

\title{Search of systematic behavior of breakup probability
in reactions with weakly bound projectiles at energies around  Coulomb barrier}
\author{V.V. Sargsyan$^{1,2}$, G.G. Adamian$^{1}$, N.V.Antonenko $^1$, W. Scheid$^3$, and  H.Q. Zhang$^4$}
\affiliation{$^{1}$Joint Institute for Nuclear Research, 141980 Dubna, Russia\\
$^{2}$International Center for Advanced Studies, Yerevan State University,  0025 Yerevan, Armenia\\
$^{3}$Institut f\"ur Theoretische Physik der Justus--Liebig--Universit\"at, D--35392 Giessen, Germany\\
$^{4}$China Institute of Atomic Energy, Post Office Box 275, Beijing 102413,  China}
\date{\today}

\begin{abstract}
Comparing  the capture cross sections  calculated  without the breakup effect
and experimental complete fusion cross sections, the breakup  was analyzed
in reactions with weakly bound projectiles $^{6,7,9}$Li, $^{9,11}$Be, and  $^{6,8}$He.
A trend of a systematic
behavior for the complete fusion suppression as a function of the target charge
and bombarding energy is not achieved.
The  quasielastic  backscattering  is suggested to be an useful
tool to study  the behavior of the breakup probability in reactions with weakly bound projectiles.

\end{abstract}


\maketitle

\section{Introduction}

In recent years, many efforts have been made to understand the  effect of  breakup of weakly
bound nuclei during the fusion reaction  in  very asymmetric reactions where
the capture cross section is equal to the complete fusion cross
section~\cite{Bertulani,PRSGomes1,Alamanos,PRSGomes2,PRSGomes3,PRSGomes4,Rafi,Alinka,PRSGomes5}.
The light radioactive nuclei, especially halo nuclei,
such as $^6$He, $^8$B, $^{11}$Be, and  the stable nuclei $^{6,7}$Li and $^{9}$Be
are weakly bounded, hence there is a chance of the breakup in the colliding process.
By performing a comparison of fusion data with theoretical predictions
which do not take into account the dynamic breakup plus transfer channel effects,
it has been shown~\cite{PRSGomes2,PRSGomes3,PRSGomes4,PRSGomes5}, that
for energies %not too much above the barrier
from about $1.1V_b$ to $1.5V_b$ ($V_b$ is the height of the Coulomb barrier)
complete fusion in the reactions $^{6,7}$Li+$^{208}$Pb,$^{209}$Bi and
$^{9}$Be+$^{89}$Y,$^{124}$Sn,$^{208}$Pb,$^{209}$Bi
is suppressed by about 30\%.
However, the $^{9}$Be+$^{144}$Sm data is out of the systematics,
showing a much smaller suppression of about 15\%.
The total fusion (incomplete fusion + sequential complete fusion + complete fusion) cross section
for the same projectiles on
targets of any mass, including $^{9}$Be + $^{27}$Al,$^{64}$Zn, does not seem to be affected by the dynamic
breakup and transfer effects~\cite{PRSGomes4,PRSGomes5}. As the charge of
the target decreases, one expects that the Coulomb breakup
becomes weaker, and consequently the complete fusion suppression and
incomplete fusion probability decrease.
The lack of a clear systematic behavior of the complete fusion suppression
as a function of the target charge was explained in Ref.~\cite{PRSGomes5} by different effects
of the transfer channels on the
complete fusion and by some problems with the experimental data analysis.

In the present article we try to reveal a systematic behavior of the complete fusion suppression
as a function of the target charge $Z_T$ and colliding energy $E_{\rm c.m.}$
by using the quantum diffusion approach~\cite{EPJSub,EPJSub1}
and by comparing the calculated capture cross sections in the absence of breakup
%, which do not take breakup into account,
with the  experimental complete and total fusion cross sections.
%  for systems for which the data are available.
The effects of deformation and neutron transfer on the complete fusion
 are taken into consideration.

\section{Model}
In the quantum diffusion approach~\cite{EPJSub,EPJSub1}
the collision of  nuclei is described with
a single relevant collective variable: the relative distance  between
the colliding nuclei. This approach takes into account  the fluctuation and dissipation effects in
collisions of heavy ions which model the coupling with various channels
(for example, coupling of the relative motion with low-lying collective modes
such as dynamical quadrupole and octupole modes of the target and projectile nuclei ~\cite{Ayik333}).
We like to mention that many quantum-mechanical and non-Markovian effects accompanying
the passage through the potential barrier are considered in our
formalism~\cite{EPJSub,EPJSub1,PRCPOP}.
The  nuclear deformation effects
are taken into account through the dependence of the nucleus-nucleus potential
on the deformations and mutual orientations of the colliding nuclei.
To calculate the nucleus-nucleus interaction potential $V(R)$,
we use the procedure presented in Refs.~\cite{EPJSub,EPJSub1}.
For the nuclear part of the nucleus-nucleus potential, a double-folding formalism with
a Skyrme-type density-dependent effective nucleon-nucleon interaction is used.
Within this approach many heavy-ion capture
reactions with stable and radioactive beams
at energies above and well below the Coulomb barrier have been
successfully described~\cite{EPJSub,EPJSub1,PRCPOP}.
One should note that other diffusion models, which  include the quantum statistical effects,
were also proposed in~\cite{Hofman}.

%Since the transfer of protons is shielded by the Coulomb barrier, it occurs when
%two nuclei almost touch each other~\cite{obzor},
%i.e. after a capture. Thus, the proton transfer can be disregarded
%in the calculations of  capture cross sections.
%Following the hypothesis of Ref.~\cite{Broglia},
We assume that the sub-barrier capture  mainly  depends  on the optimal one-neutron ($Q_{1n}>Q_{2n}$) or
two-neutron ($Q_{2n}>Q_{1n}$) transfer with a  positive  $Q$-value.
% among all possible transfer channels.
Our assumption is that, just before the projectile is captured by the target-nucleus
(just before the crossing of the Coulomb barrier) which is a slow process,
the  transfer  occurs   that can lead to the
population of the first excited collective state in the recipient nucleus~\cite{SSzilner}.
So, the motion to the
$N/Z$ equilibrium starts in the system before the capture
because it is energetically favorable in the dinuclear system in the vicinity of the Coulomb barrier.
For the reactions under consideration,
the average change of mass asymmetry is connected to the one- or two-neutron
transfer.
Since after the transfer the mass numbers, the isotopic composition and  the deformation parameters
of the interacting nuclei, and, correspondingly, the height $V_b=V(R_b)$
%($R_b$ is the position of the Coulomb barrier) are changed,
and shape of the Coulomb barrier are changed,
one can expect an enhancement or suppression of the capture.
%If  after the neutron transfer the deformation of the interacting nuclei increases (decreases),
%the capture probability increases (decreases).
When the isotopic dependence of the nucleus-nucleus interaction potential is weak and
the deformations of the interacting nuclei after the transfer have not changed,
there is no effect of the neutron transfer on the capture cross section.
This scenario was verified in the description of many reactions in Ref.~\cite{EPJSub1}.

\section{Results of calculations}

All calculated results are obtained with the same set of parameters
as in Ref.~\cite{EPJSub}.
% and are rather insensitive to the reasonable variation of them~\cite{EPJSub,EPJSub1}.
We use the friction coefficient in the relative distance coordinate
%$\hbar\lambda$=2 MeV
which  is close to that calculated within the mean field approaches~\cite{obzor}.
%The heights of the calculated Coulomb barriers $V_b=V(R_b)$
%($R_b$ is the position of the Coulomb barrier)
%are adjusted to the experimental data for the fusion or capture cross sections.
%The parameters of the nucleus-nucleus interaction potential $V(R)$
%are adjusted to describe the experimental
%data at energies above the Coulomb barrier corresponding to spherical nuclei.
The absolute values of the quadrupole deformation parameters $\beta_2$
of even-even deformed nuclei are taken from Ref.~\cite{Ram}.
For the  nuclei deformed in the
ground state, the $\beta_2$ in the first excited collective state is similar
to the $\beta_2$ in the ground state.
For the quadruple deformation parameter of an odd nucleus, we choose the maximal value of  the
deformation parameters of neighboring even-even nuclei.
% (for example, $\beta_2$($^{231}$Th)=$\beta_2$($^{233}$Th)=$\beta_2$($^{232}$Th)=0.2608).
%In Ref.~\cite{Ram} the quadrupole
%deformation parameters $\beta_2$ are given for the first excited
%2$^{+}$ states of nuclei. For the  nuclei deformed in the
%ground state, the $\beta_2$ in 2$^{+}$ state is similar
%to the $\beta_2$ in the ground state and we use $\beta_2$
%from Ref.~\cite{Ram} in the calculations.
For the double magic  and neighboring nuclei,
in the ground state we set $\beta_2=0$.
There are  uncertainties in the definition of the values of $\beta_2$
in  light-mass nuclei. However, these uncertainties weakly influence the
capture cross sections in the asymmetric reactions treated. In the calculations for light nuclei we use $\beta_2$
from Ref.~\cite{Kanada}.
%quadrupole deformation parameters of these  nuclei from a comparison
%of the calculated capture cross sections with the existing experimental data.
% (Figs.~5-8) ($Q_{2n}$ are negative or close to zero).
%The best case is
%when the projectile or target is the spherical double magic nucleus and
%there are no neutron transfer channels with  positive $Q$-values.
%There are the uncertainties in the definition of the deformation of the light nucleus
%and we have no practically experimental deformation parameters for the odd-even, even-odd, and
%odd-odd nuclei.
%By describing the
%$^{12}$C+$^{208}$Pb  reaction (Fig.~1),
%$^{18}$O+$^{208}$Pb (Fig.~2),
%$^{32,36}$S+$^{90}$Zr,
%$^{34}$S+$^{168}$Er,
%$^{36}$S+$^{90,96}$Zr,
%and $^{40}$Ca+$^{90}$Zr,
%where there are no neutron transfer channels with positive $Q$-values,
%we extract   the ground-state
%quadrupole deformation parameters $\beta_2=-0.3$,  $\beta_2=0.1$, $\beta_2=0.312$,
%$\beta_2=0.1$, and  $\beta_2=0$
%for the nuclei $^{12}$C,
%$^{18}$O, $^{32}$S
%$^{34}$S, and $^{36}$S, respectively~\cite{EPJSub1} which are used in our calculations.
%For the   $^{32}$S nucleus, the extracted $\beta_2$ is equal to the
%experimental one from Ref.~\cite{Ram}.

\subsection{Breakup probabilities}
In Figs.~1-13 we compare the calculated $\sigma_c^{th}$ capture cross sections
with the experimental $\sigma_{fus}^{exp}$ complete and total fusion cross sections
in the reactions induced by projectiles
$^{9}$Be, $^{10,11}$B, $^{6,7,9}$Li, $^{4,6,8}$He
\cite{B11Bi209,B11Tb159,Be9Bi209,Be9Pb208,Be9Sm144,Be9Sn124,Be9Y89,Be9Zn64,Be9Al27,Li7Al27,Li7Au197,Li6Bi209,Li6Pb208,Li6Pt198,Li6Sm144,Li6Zn64,Li7Ho165,He4Bi211,He4Zn64,Li9Zn70,He4Au197,Li9Pb208}.
The difference between the capture cross section and
the complete fusion cross section can be ascribed
to the breakup effect.
Comparing $\sigma_c^{th}$ and $\sigma_{fus}^{exp}$, one can estimate the breakup probability
\begin{eqnarray}
P_{\rm BU}=1-\sigma_{fus}^{exp}/\sigma_c^{th}.
\end{eqnarray}
If at some energy $\sigma_{fus}^{exp}>\sigma_c^{th}$, the values
of $\sigma_c^{th}$ was normalized so to have $P_{\rm BU}\ge 0$ at any energy.

Note that $\sigma_{fus}^{exp}=\sigma_{fus}^{noBU}+\sigma_{fus}^{BU}$
contains the contribution from two processes: the direct fusion of the
projectile with the target ($\sigma_{fus}^{noBU}$), and the breakup of the
projectile followed by the fusion of the two projectile fragments with
the target ($\sigma_{fus}^{BU}$). A more adequate estimate of the breakup
probability would then be:
\begin{eqnarray}
P_{\rm BU}=1-\sigma_{fus}^{noBU}/\sigma_c^{th},
\end{eqnarray}
which
leads to larger values of $P_{\rm BU}$ than the expression employed by us.
However, the ratio between $\sigma_{fus}^{noBU}$ and $\sigma_{fus}^{BU}$
cannot be measured experimentally, but can be estimated with the approach
suggested in Refs.~\cite{Maximka,PLATYPUS}. The parameters of the potential
are taken to fit the height of the Coulomb barrier obtained in our calculations.
The parameters of the breakup function \cite{Maximka} are set to describe the value of
$\sigma_{fus}^{exp}$. As shown in Ref.~\cite{Maximka} and in our calculations, in
the $^8$Be+$^{208}$Pb reaction the fraction of
$\sigma_{fus}^{BU}$ in $\sigma_{fus}^{exp}$ does not exceed few percents
at $E_{\rm c.m.}-V_b<$4 MeV. This fraction rapidly increases and reaches about 12--20\%, depending on the reaction, at
$E_{\rm c.m.}-V_b\approx$10 MeV. Because we are mainly interested in
the energies near and below the barrier, the estimated $\sigma_{fus}^{BU}$ does not exceed 20\% of
$\sigma_{fus}^{exp}$ at $E_{\rm c.m.}-V_b <$10 MeV. The results for $P_{\rm BU}$ are presented
taking $\sigma_{fus}^{noBU}$ into account in Eq.~(2).

As seen in Figs.~14 and 15, at energies above the Coulomb barriers the values of $P_{\rm BU}$ vary from 0 to 84\%.
In the reactions $^9$Be+$^{144}$Sm,$^{208}$Pb,$^{209}$Bi the value of $P_{\rm BU}$ increases with charge
number of the target at $E_{\rm c.m.}-V_b>3$ MeV.
This was also noted in Ref.~\cite{PRSGomes5}. However, the reactions $^9$Be+$^{89}$Y,$^{124}$Sn are out of this
systematics. In the reactions $^6$Li+$^{144}$Sm,$^{198}$Pt,$^{209}$Bi the value of $P_{\rm BU}$ decreases with
increasing charge number of the target at $E_{\rm c.m.}-V_b>3$ MeV. While in the reactions
$^9$Be+$^{89}$Y,$^{144}$Sm,$^{208}$Pb,$^{209}$Bi the value of  $P_{\rm BU}$ has a minimum
at $E_{\rm c.m.}-V_b\approx 0$ and a maximum at $E_{\rm c.m.}-V_b\approx -(1-3)$ MeV, in the
$^9$Be+$^{124}$Sn reaction the value of $P_{\rm BU}$ steady decreases with energy. In the reactions
$^6$Li+$^{144}$Sm,$^{198}$Pt,$^{209}$Bi, $^7$Li+$^{208}$Pb,$^{209}$Bi, and $^9$Li+$^{208}$Pb there is maximum of
$P_{\rm BU}$
at $E_{\rm c.m.}-V_b\approx -(0-1)$ MeV. However, in the reactions $^6$Li+$^{208}$Pb and $^7$Li+$^{165}$Ho $P_{\rm BU}$
has a minima $E_{\rm c.m.}-V_b\approx 2$ MeV and no maxima at $E_{\rm c.m.}-V_b\approx 0$.
For $^9$Be, the breakup threshold is slightly larger than for $^6$Li.
Therefore, we can not explain a larger breakup probability
at smaller $E_{\rm c.m.}-V_b$ in the case of $^9$Be.

In Figs.~1-13 we also show the calculated
capture cross sections normalized by some factors to
obtain a rather good agreement between the experimental and theoretical results.
These average normalization factors are
0.7, 0.75, 0.9, 0.64, 0.7, 1, 0.9
for the reactions $^9$Be+$^{209}$Bi,$^{208}$Pb,$^{144}$Sm,$^{124}$Sn,$^{89}$Y,$^{64}$Zn,$^{27}$Al,
respectively,
0.52, 0.5, 0.5,  0.42, 0.65
for the reactions  $^6$Li+$^{209}$Bi,$^{208}$Pb,$^{198}$Pt,$^{144}$Sm,$^{64}$Zn, respectively,
0.6, 0.7, 0.65, 0.6, 0.65, 0.75
for the reactions  $^7$Li+$^{209}$Bi,$^{197}$Au,$^{165}$Ho,$^{159}$Tb,$^{64}$Zn,$^{27}$Al,  respectively.
For the reactions
$^9$Li+$^{208}$Pb (Fig.~10),
$^6$He+$^{209}$Bi (Fig.~11), $^6$He+$^{64}$Zn (Fig.~11),
$^{6}$He+$^{197}$Au (Fig.~12), $^{8}$He+$^{197}$Au  (Fig.~12),
$^{11}$B+$^{209}$Bi (Fig.~13) and $^{11}$B+$^{159}$Tb (Fig.~13),
a  satisfactory agreement between experimental fusion data and
capture cross sections can be reached with  average  normalization
factors   0.6, 0.68, 0.4, 0.8, 0.7, 0.82, and 0.95, respectively. Note that these  average
normalization  factors  do not depend on $E_{\rm c.m.}$. With the $^{9}$Be projectile
we  obtain the complete fusion suppressions  similar to those  reported in Refs.~\cite{PRSGomes4,PRSGomes5}.
For lighter targets, when the Coulomb breakup becomes weaker, one expects that the suppression of
complete fusion becomes smaller than for heavy targets.

An expected behavior for complete fusion suppression is that
fusion probability increases with decreasing $Z_T$. However, one can observe deviations from this rule.
In the reactions $^{9}$Be+$^{124}$Sn,$^{89}$Y the data show quite larger complete fusion suppression (30--36)\%.
For the reactions induced by a $^{6}$Li projectile,
one can see that  the fusion suppression is nearly independent of $Z_T$.
The replacement of $^{7}$Li by $^{6}$Li in the reactions in Fig.~9 almost does not change the
experimental~\cite{Li7Au197,Li7Al27} and calculated data.
The Coulomb fields for very light systems $^9$Be+$^{27}$Al,  and $^6$He,$^7$Li+$^{64}$Zn
are not strong enough to produce an appreciable
breakup. It is not realistic that the  fusion suppression
in the  $^9$Be+$^{64}$Zn reaction is smaller than the one in the $^9$Be+$^{27}$Al reaction
or the suppressions of fusion  coincide in the reactions $^{4,6}$He+$^{64}$Zn (Fig.~11) with stable
and exotic projectiles. Note that the experimental data for the reactions
$^{6,7}$Li+$^{27}$Al,$^{64}$Zn and $^{9}$Be+$^{27}$Al,$^{64,70}$Zn  are for the total fusion.
In general, the total fusion does not seem to be affected by breakup~\cite{PRSGomes4,PRSGomes5}.

So, there is a lack of a systematic behavior of the complete fusion
suppression for the systems treated.
The possible explanation of it is that there are probably some problems with the data analysis
%, for example,  if there is an inaccuracy in data normalization
which were earlier noted in Refs.~\cite{PRSGomes4,PRSGomes5} from the point of view
of a universal fusion function representation.
%For instance, from our analysis follows that the suppressions of
%complete fusion in the reactions $^{4,6}$He+$^{64}$Zn with the stable $^{4}$He
%and  $^{6}$He exotic projectiles almost coincide which is not physical result.
It could be also that at energies near the Coulomb barrier
the characteristic time of the breakup is larger than
the characteristic time of the capture process and influences the complete fusion.
For the reactions $^{6,7}$Li+$^{208}$Pb,
the  characteristic  times of the prompt and delayed breakup  were studied recently in Ref.~\cite{Luong}.

The large positive $Q_{2n}$-value in the $^9$Li+$^{208}$Pb reaction~\cite{Li9Pb208}
gives a possibility of a two-neutron transfer before the capture. However,  the capture cross sections
calculated with and without neutron transfer are very close to each other because the effect of neutron transfer
is rather weak in asymmetric reactions~\cite{EPJSub,EPJSub1}.
The calculated capture cross sections normalized by a factor of
0.6 are shown by the dotted line in the lower part of Fig.~10.
In the upper part of  Fig.~10, the predicted capture cross
sections for the reaction $^{11}$Li+$^{208}$Pb are shown.

%At last in Fig.~16 the capture cross sections for the reactions
%$^{9}$Be+$^{182}$W and $^{6}$Li+$^{184}$W are predicted.

%We assume, that the carefully measurements of the fusion cross section
%in these reactions will coincide with the predicted capture cross section.

%Also a large positive $Q$ value is present in the reaction $^6$He+$^{209}$Bi (Fig.~13).
%If one include in the calculation the possibility of two neutron transfer, then
%a good agreement between the fusion data and capture cross section is reached with the coefficient 0.68.
%In Fig.~13 also presented  the comparison for the  $^4$He+$^{64}$Zn and $^6$He+$^{64}$Zn reactions.
%The coefficients for both reactions is 0.4.
%The situation is not so clear in the case of the reactions $^9$Li+$^{70}$Zn
%and $^{4,6,8}$He+$^{197}$Au reactions (Fig.~14).
%In the case of $^9$Li+$^{70}$Zn reaction there is a possibility of two neutron transfer.
%However, the capture cross section in both cases practically the same, and
%are in disagreement with experimental fusion data.
%In the case of $^{6}$He+$^{197}$Au and $^{8}$He+$^{197}$Au reactions
%we take into account the neutron transfers $^{4}$He+$^{199}$Au and $^{6}$He+$^{199}$Au,
%respectively. The experimental error bars in the reactions $^{6}$He+$^{197}$Au
%does not give one to make conclusion about contribution of breakup.
%However, for the reactions  $^{4}$He+$^{197}$Au and $^{8}$He+$^{197}$Au
%(with including two neutron transfer channel) a  satisfactory agreement
%between fusion data and capture cross section could be reached with normalizing coefficients 0.7 and 0.8.

\subsection{Quasielastic backscattering - tool for search of  breakup process
in reactions with weakly bound projectiles}
The lack of a clear systematic behavior of the complete fusion suppression
as a function of the target charge  requires new additional
experimental and theoretical studies.
The quasielastic backscattering
has been used~\cite{Timmers,Zhang,Sinha,Piasecki}
as an alternative to investigate fusion (capture) barrier
distributions, since this process is complementary to fusion.
%The sum of elastic and inelastic scattering is much easier to
%measure than fusion, and its barrier distribution is derived by
%the first derivative of its relative cross section to the Rutherford
%cross section with respect to the center-of-mass energy.
Since
the quasielastic experiment is usually not as complex as the capture (fusion) and breakup
measurements, they are well suited to survey the breakup probability.
There is a direct relationship between the capture,
the quasielastic scattering  and the breakup processes, since any loss from
the quasielastic and breakup channel contributes directly to capture (the conservation of the reaction flux):
\begin{eqnarray}
P_{qe}(E_{\rm c.m.},J)+P_{cap}(E_{\rm c.m.},J)+P_{BU}(E_{\rm c.m.},J)=1,
\end{eqnarray}
where
$P_{qe}$  is the reflection quasielastic probability, $P_{BU}$ is the breakup (reflection) probability, and
$P_{cap}$ is the capture (transmission) probability.
The quasielastic scattering is the sum of
all direct reactions, which include elastic, inelastic, and transfer processes.
%
%It has been proposed that to obtain
%the interaction barrier, quasielastic scattering should be measured
%at backward angles of nearly 180 degrees, where head-on collision
%is dominant~\cite{Timmers,Zhang,Sinha,Piasecki}.
%
Equation~(3) can be rewritten as
\begin{eqnarray}
\frac{P_{qe}(E_{\rm c.m.},J)}{1-P_{BU}(E_{\rm c.m.},J)}+\frac{P_{cap}(E_{\rm c.m.},J)}{1-P_{BU}(E_{\rm c.m.},J)}
=P_{qe}^{nBU}(E_{\rm c.m.},J)+P_{cap}^{nBU}(E_{\rm c.m.},J)=1,
\end{eqnarray}
where
$$P_{qe}^{nBU}(E_{\rm c.m.},J)=\frac{P_{qe}(E_{\rm c.m.},J)}{1-P_{BU}(E_{\rm c.m.},J)}$$
and
$$P_{cap}^{nBU}(E_{\rm c.m.},J)=\frac{P_{cap}(E_{\rm c.m.},J)}{1-P_{BU}(E_{\rm c.m.},J)}$$
are the quasielastic and  capture  probabilities, respectively,  in the absence of the
breakup process. From these expressions we obtain the useful formulas
\begin{eqnarray}
\frac{P_{qe}(E_{\rm c.m.},J)}{P_{cap}(E_{\rm c.m.},J)}=\frac{P_{qe}^{nBU}(E_{\rm c.m.},J)}{P_{cap}^{nBU}(E_{\rm c.m.},J)}
=\frac{P_{qe}^{nBU}(E_{\rm c.m.},J)}{1-P_{qe}^{nBU}(E_{\rm c.m.},J)}=a.
\end{eqnarray}
Using  Eqs.~(3) and (5), we obtain the relationship between breakup and quasielastic processes:
\begin{eqnarray}
P_{BU}(E_{\rm c.m.},J)&=&1-[P_{qe}(E_{\rm c.m.},J)+P_{cap}(E_{\rm c.m.},J)]=1-P_{qe}(E_{\rm c.m.},J)[1+1/a]  \nonumber\\
&=&
1-P_{qe}(E_{\rm c.m.},J)/P_{qe}^{nBU}(E_{\rm c.m.},J).
\end{eqnarray}
The last equation is one of important results of the present paper.
Analogously one can find other expression
\begin{eqnarray}
P_{BU}(E_{\rm c.m.},J)=
1-P_{cap}(E_{\rm c.m.},J)/P_{cap}^{nBU}(E_{\rm c.m.},J),
\end{eqnarray}
which relates the breakup and capture processes.

The reflection quasielastic probability
\begin{eqnarray}
P_{qe}(E_{\rm c.m.},J=0)=d\sigma_{qe}/d\sigma_{Ru}
\end{eqnarray}
for bombarding energy
$E_{\rm c.m.}$ and
angular momentum $J=0$ is given by the ratio of
the quasielastic differential cross section $\sigma_{qe}$  and
Rutherford differential cross section $\sigma_{Ru}$
at 180 degrees~\cite{Timmers,Zhang,Sonzogni,Sinha,Piasecki}.
Employing Eqs.~(6), (8),  and the  experimental quasielastic backscattering data
with toughly
and
weakly bound isotopes-projectiles and the same compound nucleus,
one can extract the breakup probability of the exotic nucleus.
For example, using Eq.~(6) at $J=0$ and the experimental $P_{qe}^{nBU}$[$^{4}$He+$^{208}$Pb] of the
$^{4}$He+$^{208}$Pb reaction with toughly bound nuclei (without breakup) and $P_{qe}$[$^{6}$He+$^{206}$Pb]
of the $^{6}$He+$^{206}$Pb reaction with weakly bound projectile (with breakup), and taking into consideration
$V_b$($^{4}$He+$^{208}$Pb)$\approx V_b$($^{6}$He+$^{206}$Pb)
for the very asymmetric systems,
one can extract the breakup probability of the $^{6}$He:
\begin{eqnarray}
P_{BU}(E_{\rm c.m.},J=0)=
1-\frac{P_{qe}(E_{\rm c.m.},J=0)[^{6}He+^{206}Pb]}{P_{qe}^{nBU}(E_{\rm c.m.},J=0)[^{4}He+^{208}Pb]}.
\end{eqnarray}
Comparing the experimental quasielastic backscattering
cross sections in the presence and absence  of breakup
data in the reaction pairs
$^{6}$He+$^{68}$Zn and $^{4}$He+$^{70}$Zn,
$^{6}$He+$^{122}$Sn and $^{4}$He+$^{124}$Sn,
$^{6}$He+$^{236}$U and $^{4}$He+$^{238}$U,
$^{8}$He+$^{204}$Pb and $^{4}$He+$^{208}$Pb,
$^{9}$Be+$^{208}$Pb and  $^{10}$Be+$^{207}$Pb,
$^{11}$Be+$^{206}$Pb and  $^{10}$Be+$^{207}$Pb,
$^{8}$B+$^{208}$Pb and $^{10}$B+$^{206}$Pb,
$^{8}$B+$^{207}$Pb and $^{11}$B+$^{204}$Pb,
$^{9}$B+$^{208}$Pb and $^{11}$B+$^{206}$Pb,
$^{15}$C+$^{207}$Pb and $^{14}$C+$^{208}$Pb,
$^{17}$F+$^{206}$Pb and $^{19}$F+$^{208}$Pb
leading to the same corresponding compound nuclei,
one can analyze  the role of the breakup channels
in the reactions with
the light weakly bound projectiles $^{6,8}$He, $^{9,11}$Be,
 $^{8,9}$B, $^{15}$C, and  $^{17}$F  at near and below barrier
energies.
One concludes that
the quasielastic technique could be a very important tool in breakup research.
We propose to extract the breakup probability
directly from the quasielastic cross sections of systems mentioned above.
%\section{Summary}

\section{Summary}
Comparing the calculated capture cross sections in the absence of breakup
data and experimental complete fusion
data, we analyzed the role of the breakup channels in the reactions with
the light projectiles $^{9}$Be, $^{6,7,9}$Li and $^{6,8}$He at near-barrier
energies. Within the quantum diffusion approach the neutron transfer and deformation effects
were taken into account. Analyzing the extracted breakup probabilities, we
showed that there are no systematic trends of breakup in the  reactions studied.
Moreover, for some system with larger (smaller) $Z_T$ we found the contribution of breakup to be smaller (larger).
Almost for all reactions  considered we obtained  a satisfactory agreement between calculated capture cross section
and experimental fusion data, if the calculated capture cross section
or the experimental fusion data are renormalized by some average factor which does not depend on the bombarding
energy.
Note that our conclusions  coincide with those of Refs.~\cite{PRSGomes4,PRSGomes5}, where
the universal fusion function formalism was applied for the analysis of experimental data.
One needs to measure directly   the breakup process in different systems,
especially light ones, to understand the role of the Coulomb
breakup in the complete fusion process.
%The first steps in this direction were done in Ref.~\cite{Rafi}.
%The new measurements and calculations  are also very important
%in such studies.
The other important subject to be investigated both experimental
and theoretically is the characteristic
time of the breakup.
The first steps in these directions were done
in Refs.~\cite{Rafi,Gomesnew,Luong}.

As shown, one no needs to measure directly   the breakup process
in different systems,
especially light ones, to understand the role of the
breakup in the  capture (complete fusion) process.
Employing the experimental quasielastic backscattering data
with weakly and toughly bound isotopes of light nucleus
and Eq.~(6), the dependence of
breakup probability on $E_{\rm c.m.}$ can be extracted for the systems suggested.
Analyzing the extracted breakup probabilities, one can indirectly
study the trends of breakup in the different reactions at energies
near and below Coulomb barrier.
%and understand the role of the
%breakup in the  capture (complete fusion) process.
%

\section{acknowledgements}
We thank P.R.S. Gomes for fruitful discussions and  useful suggestions.

%\newpage
\begin{figure}
\caption{The calculated capture cross sections vs $E_{\rm c.m.}$ for the reactions
$^{9}$Be+$^{209}$Bi and $^{9}$Be+$^{208}$Pb (solid lines).
The experimental data (squares) are  from Refs.~\protect\cite{Be9Bi209,Be9Pb208}.
The calculated capture cross sections  normalized by factors 0.7 and 0.75 for
the  reactions $^{9}$Be+$^{209}$Bi and $^{9}$Be+$^{208}$Pb, respectively, are presented by dotted lines.
%The following  quadrupole deformation parameters  are used:
%$\beta_{2}$($^{9}$Be)=0.9, $\beta_{2}$($^{209}$Bi)=0.05~\protect\cite{Ram} and $\beta_{2}$($^{208}$Pb)=0.
}
\label{1_fig}
\end{figure}

\begin{figure}
\caption{(Color online) The calculated capture cross sections vs $E_{\rm c.m.}$ for the reactions
$^{9}$Be+$^{144}$Sm and $^{9}$Be+$^{124}$Sn  (solid lines).
The experimental data (squares) are  from Refs.~\protect\cite{Be9Sm144,Be9Sn124}.
The experimental total fusion data~\protect\cite{Be9Sm144} for the $^{9}$Be+$^{144}$Sm reaction
are shown by  open circles.
The calculated capture cross sections normalized by factors  0.9 and 0.64 for
the  reactions $^{9}$Be+$^{144}$Sm and $^{9}$Be+$^{124}$Sn, respectively, are presented by dotted lines.
%The following  quadrupole deformation parameters  are used:
%$\beta_{2}$($^{9}$Be)=0.9, $\beta_{2}$($^{144}$Sm)=0.05~\protect\cite{Ram} and
%$\beta_{2}$($^{124}$Sn)=0.0953~\protect\cite{Ram}.
}
\label{2_fig}
\end{figure}

\begin{figure}
\caption{The calculated capture cross sections vs $E_{\rm c.m.}$ for the reactions
$^{9}$Be+$^{89}$Y  and $^{9}$Be+$^{64}$Zn  (solid lines).
The experimental data (squares) are from  Refs.~\protect\cite{Be9Y89,Be9Zn64}.
The calculated capture cross sections  normalized by a factor  0.7 for
the $^{9}$Be+$^{89}$Y  reaction  are presented by a dotted line.
%The following  quadrupole deformation parameters  are used:
%$\beta_{2}$($^{9}$Be)=0.9, $\beta_{2}$($^{89}$Y)=0.05~\protect\cite{Ram} and
%$\beta_{2}$($^{64}$Zn)=0.242~\protect\cite{Ram}.
}
\label{3_fig}
\end{figure}

\begin{figure}
\caption{The calculated capture cross sections vs $E_{\rm c.m.}$ for the reaction
$^{9}$Be+$^{27}$Al (solid line).
The experimental data (squares) are  from  Ref.~\protect\cite{Be9Al27}.
The calculated capture cross sections  normalized by a factor  0.9 for
the $^{9}$Be+$^{27}$Al reaction are presented by a dotted line.
%The dashed line is the solid line shifted by 1.8 MeV to the higher energies.
%The following  quadrupole deformation parameters  are used:
%$\beta_{2}$($^{9}$Be)=0.9, $\beta_{2}$($^{238}$U)=0.2863 and $\beta_{2}$($^{27}$Al)=0.3.
}
\label{4_fig}
\end{figure}

\begin{figure}
\caption{The calculated capture cross sections vs $E_{\rm c.m.}$ for the reactions
$^{6}$Li+$^{209}$Bi and $^{6}$Li+$^{208}$Pb (solid lines).
The experimental data (squares) are from  Refs.~\protect\cite{Li6Bi209,Li6Pb208}.
The calculated capture cross sections normalized by factors  0.52 and 0.5 for
the  reactions $^{6}$Li+$^{209}$Bi and $^{6}$Li+$^{208}$Pb, respectively, are presented by dotted lines.
%The following  quadrupole deformation parameters  are used:
%$\beta_{2}$($^{6}$Li)=0.4, $\beta_{2}$($^{209}$Bi)=0.05~\protect\cite{Ram} and $\beta_{2}$($^{208}$Pb)=0.
}
\label{5_fig}
\end{figure}

\begin{figure}
\caption{The calculated capture cross sections vs $E_{\rm c.m.}$ for the reactions
$^{6}$Li+$^{198}$Pt and $^{6}$Li+$^{144}$Sm (solid lines).
The experimental data (squares) are  from  Refs.~\protect\cite{Li6Pt198,Li6Sm144}.
The calculated capture cross sections normalized by factors 0.5 and 0.42 for
the  reactions $^{6}$Li+$^{198}$Pt and $^{6}$Li+$^{144}$Sm, respectively, are presented by dotted lines.
%The following  quadrupole deformation parameters  are used:
%$\beta_{2}$($^{6}$Li)=0.4, $\beta_{2}$($^{198}$Pt)=0.1141~\protect\cite{Ram} and $\beta_{2}$($^{144}$Sm)=0.
}
\label{6_fig}
\end{figure}

\begin{figure}
\caption{The calculated capture cross sections vs $E_{\rm c.m.}$ for the reactions
$^{7}$Li+$^{209}$Bi and $^{7}$Li+$^{64}$Zn (solid lines).
The calculated results for the reactions $^{6,7}$Li+$^{64}$Zn  almost coincide.
The experimental data (squares) are  from  Refs.~\protect\cite{Li6Bi209,Be9Zn64}.
The experimental data for the reactions $^{7}$Li+$^{64}$Zn (squares) and
$^{6}$Li+$^{64}$Zn (circles and stars) are  from  Refs.~\protect\cite{Be9Zn64,Li6Zn64}.
The calculated capture cross sections normalized by factors  0.6 and 0.65 for
the  reactions $^{7}$Li+$^{209}$Bi and $^{7}$Li+$^{64}$Zn, respectively, are presented by dotted lines.
%The following  quadrupole deformation parameters  are used:
%$\beta_{2}$($^{7}$Li)=0.4, $\beta_{2}$($^{209}$Bi)=0.05~\protect\cite{Ram}
%and $\beta_{2}$($^{64}$Zn)=0.242~\protect\cite{Ram}.
}
\label{7_fig}
\end{figure}

\begin{figure}
\caption{The calculated capture cross sections vs $E_{\rm c.m.}$ for the reactions
$^{7}$Li+$^{165}$Ho and $^{7}$Li+$^{159}$Tb (solid lines).
The experimental data (squares) are  from  Refs.~\protect\cite{Li7Ho165,B11Tb159}.
The experimental total fusion data~\protect\cite{Li7Ho165,B11Tb159} are shown by the solid triangles.
The calculated capture cross sections normalized by factors  0.65 and 0.6 for
the  reactions $^{7}$Li+$^{165}$Ho and $^{7}$Li+$^{159}$Tb, respectively, are presented by dotted lines.
%The following  quadrupole deformation parameters  are used:
%$\beta_{2}$($^{7}$Li)=0.4, $\beta_{2}$($^{165}$Ho)=0.3481~\protect\cite{Ram} and
%$\beta_{2}$($^{159}$Tb)=0.3484~\protect\cite{Ram}.
}
\label{8_fig}
\end{figure}

\begin{figure}
\caption{The calculated capture cross sections vs $E_{\rm c.m.}$ for the reactions
$^{7}$Li+$^{197}$Au and $^{7}$Li+$^{27}$Al (solid lines).
The experimental data (squares) are  from  Refs.~\protect\cite{Li7Au197,Li7Al27}.
%The experimental data~\protect\cite{Li7Au197,Li7Al27}
%for the reactions $^{6}$Li+$^{27}$Al,$^{197}$Au are marked by circles.
The calculated capture cross sections normalized by factors  0.7 and
0.75 for the  reactions $^{7}$Li+$^{197}$Au and $^{7}$Li+$^{27}$Al, respectively, are presented by dotted lines.
%The following  quadrupole deformation parameters  are used:
%$\beta_{2}$($^{7}$Li)=0.4, $\beta_{2}$($^{197}$Au)=0.1296~\protect\cite{Ram} and $\beta_{2}$($^{27}$Al)=0.3.
}
\label{9_fig}
\end{figure}

\begin{figure}
\caption{The calculated capture cross sections vs $E_{\rm c.m.}$ for the reactions $^{11}$Li+$^{208}$Pb and
$^{9}$Li+$^{208}$Pb (solid lines).
The experimental data (squares) are  from  Ref.~\protect\cite{Li9Pb208}.
The calculated capture cross sections normalized by a factor  0.6 for
the $^{9}$Li+$^{208}$Pb reaction are presented by a dotted line.
%The calculations without neutron transfer are shown by dashed lines.
%The following  quadrupole deformation parameters  are used:
%$\beta_{2}$($^{9}$Li)=0.2, $\beta_{2}$($^{7}$Li)=0.4, $\beta_{2}$($^{210}$Pb)=$\beta_{2}$($^{208}$Pb)=0.
}
\label{10_fig}
\end{figure}

\begin{figure}
\caption{The calculated capture cross sections vs $E_{\rm c.m.}$ for the indicated reactions
$^{6}$He+$^{209}$Bi (solid line), $^{6}$He+$^{64}$Zn (solid line), and
$^{4}$He+$^{64}$Zn (dashed line).
The experimental data for the reactions $^{4}$He+$^{64}$Zn (solid squares) and
$^{6}$He+$^{64}$Zn (open squares) are  from  Refs.~\protect\cite{He4Bi211,He4Zn64}.
The calculated capture cross sections normalized by factors  0.68, 0.4, and 0.4 for
the  reactions $^{6}$He+$^{209}$Bi (dotted line), $^{6}$He+$^{64}$Zn (dotted line),
and $^{4}$He+$^{64}$Zn (dash-dotted line),
respectively, are shown. %To reproduce
%the experimental data for the $^{6}$He+$^{64}$Zn reaction, the corresponding Coulomb barrier
%is shifted by 1.3 MeV to the lower energies.
%
%The following  quadrupole deformation parameters  are used:
%$\beta_{2}$($^{4}$He)=0, $\beta_{2}$($^{211}$Bi)=0.05~\protect\cite{Ram}
%and $\beta_{2}$($^{64}$Zn)=0.242~\protect\cite{Ram}.
}
\label{11_fig}
\end{figure}

\begin{figure}
\caption{(Color online) The calculated capture cross sections vs $E_{\rm c.m.}$ for the reactions
$^{9}$Li+$^{70}$Zn (solid line),
$^{4,6}$He+$^{197}$Au (solid lines)
and
$^{8}$He+$^{197}$Au (dashed line).
The results for the reactions $^{4,6}$He+$^{197}$Au almost coincide.
The experimental data for the reactions with $^{9}$Li, $^{4}$He (solid squares), $^{6}$He (open squares), and
$^{8}$He (solid triangles) are  from  Refs.~\protect\cite{Li9Zn70,He4Au197}.
The calculated capture cross sections normalized by factors  0.8 and 0.6 for
the  reactions $^{4,6}$He+$^{197}$Au (dotted lines) and $^{8}$He+$^{197}$Au (dash-dotted line), respectively,
are shown.
%The following  quadrupole deformation parameters  are used:
%$\beta_{2}$($^{4}$He)=$\beta_{2}$($^{6}$He)=0, $\beta_{2}$($^{7}$Li)=0.4, $\beta_{2}$($^{72}$Zn)=0.253~\protect\cite{Ram},
%$\beta_{2}$($^{197}$Au)=0.1296~\protect\cite{Ram} and $\beta_{2}$($^{199}$Au)=0.1141~\protect\cite{Ram}.
}
\label{12_fig}
\end{figure}

\begin{figure}
\caption{(Color online) The calculated capture cross sections vs $E_{\rm c.m.}$ for the reactions
 $^{10,11}$B+$^{209}$Bi and $^{10,11}$B+$^{159}$Tb  (solid lines).
%The experimental data for the ... (squares) are  from  Refs.~\protect\cite{B11Bi209,B11Tb159}.
The experimental data~\protect\cite{B11Bi209,B11Tb159}
for the  reactions $^{10}$B+$^{159}$Tb,$^{209}$Bi and $^{11}$B+$^{159}$Tb,$^{209}$Bi are marked by circles
and squares, respectively.
The experimental total fusion data~\protect\cite{B11Bi209} are shown by the solid stars.
The calculated capture cross sections normalized by factors  0.82
and 0.95 for the reactions $^{11}$B+$^{209}$Bi and $^{11}$B+$^{159}$Tb, respectively, are presented by dotted lines.
%The following  quadrupole deformation parameters  are used:
%$\beta_{2}$($^{11}$B)=0.2~\protect\cite{Ram}, $\beta_{2}$($^{209}$Bi)=0.05~\protect\cite{Ram}
%and $\beta_{2}$($^{159}$Tb)=0.3484~\protect\cite{Ram}.
}
\label{13_fig}
\end{figure}

\begin{figure}
\caption{(Color online) The dependence of the extracted breakup probability $P_{BU}$
vs $E_{c.m.}-V_b$ for the indicated reactions with $^{9}$Be-projectiles in \%.
Formula (2) was used.
}
\label{14_fig}
\end{figure}

\begin{figure}
\caption{(Color online) The same as in Fig.~14, but for the indicated reactions with $^{6,7,9}$Li-projectiles.
}
\label{15_fig}
\end{figure}

%\begin{figure}
%\caption{The predicted capture cross sections versus $E_{\rm c.m.}$ for the  reactions
%$^{9}$Be+$^{182}$W and $^{8}$Li+$^{184}$W.
%%The following  quadrupole deformation parameters  are used:
%%$\beta_{2}$($^{9}$Be)=0.9, $\beta_{2}$($^{182}$W)=0.251~\protect\cite{Ram}, $\beta_{2}$($^{6}$Li)=0.4
%%and $\beta_{2}$($^{184}$W)=0.2362~\protect\cite{Ram}.
%}
%\label{16_fig}
%\end{figure}


\begin{thebibliography}{99}

\bibitem{Bertulani} C.A.~Bertulani, EPJ Web Conf. {\bf 17}, 15001 (2011);
P.R.S.~Gomes, J.~Lubian, and L.F.~Canto, EPJ Web Conf. {\bf 17}, 11003 (2011),
P.R.S.~Gomes {\it et al.}, EPJ Web Conf. (2012) in print.
%%%%%              Gomes          %%%%%%%%%%%%%%%%
\bibitem{PRSGomes1} L.F.~Canto, P.R.S.~Gomes, R.~Donangelo, and M.S.~Hussein,
Phys. Rep. {\bf 424}, 1 (2006).
\bibitem{Alamanos} N.~Keeley, R.~Raabe,  N.~Alamanos, and J.L.~Sida,
Prog. Part. Nucl. Phys. {\bf 59}, 579 (2007).
\bibitem{PRSGomes2} L.F.~Canto, P.R.S.~Gomes, J.~Lubian, L.C.~Chamon, and
E.~Crema, J. Phys. G {\bf 36}, 015109 (2009).
\bibitem{PRSGomes3} L.F.~Canto, P.R.S.~Gomes, J.~Lubian, L.C.~Chamon, and
E.~Crema, Nucl. Phys. {\bf A821}, 51 (2009).
\bibitem{PRSGomes4} P.R.S.~Gomes, J.~Lubian, and L.F.~Canto, Phys. Rev. C {\bf 79},
027606 (2009).
\bibitem{Rafi}           R.~Rafiei {\it et al.}, Phys. Rev. C {\bf 81}, 024601 (2010).
\bibitem{Alinka} P.N.~de~Faria,  R.~Lichtenth\"aler, K.C.C.~Pires, A.M.~Moro,
A.~L\'epine-Szily, V.~Guimar\~aes, D.R.~Mendes,~Jr., A.~Arazi, A.~Barioni,
V.~Morcelle, and M.C.~Morais,
 Phys. Rev. C {\bf 82},
034602 (2010); P.~Mohr, P.N.~de~Faria, R.~Lichtenth\"aler, K.C.C.~Pires,
V.~Guimar\~aes, A.~L\'epine-Szily, D.R.~Mendes,~Jr., A.~Arazi, A.~Barioni,
V.~Morcelle, and M.C.~Morais, Phys. Rev. C {\bf 82},
044606 (2010); A.~L\'epine-Szily {\it et al.}, EPJ Web Conf. (2012) in print.
\bibitem{PRSGomes5} P.R.S.~Gomes, R.~Linares, J.~Lubian, C.C.~Lopes,
E.N.~Cardozo, B.H.F.~Pereira, and I.~Padron,
Phys. Rev. C {\bf 84}, 014615 (2011).
%%%%%              Theor          %%%%%%%%%%%%%%%%
\bibitem{EPJSub}  V.V.~Sargsyan, G.G.~Adamian, N.V.~Antonenko, and W.~Scheid,
Eur. Phys. J. A {\bf 45}, 125 (2010);
V.V.~Sargsyan, G.G.~Adamian, N.V.~Antonenko,  W.~Scheid, and H.Q.~Zhang,
Eur. Phys. J. A {\bf 47}, 38 (2011); J. of Phys.: Conf. Ser. {\bf 282}, 012001 (2011);
EPJ Web Conf. {\bf 17}, 04003 (2011);
V.V.~Sargsyan, G.G.~Adamian, N.V.~Antonenko,  W.~Scheid, C.J.~Lin,
and H.Q.~Zhang, Phys. Rev. C {\bf 85}, 017603 (2012);
 {\it ibid}  {\bf 85}, 037602 (2012).
%\bibitem{EPJSub1}        V.V.~Sargsyan~{\it et al.},
%%, G.G.~Adamian, N.V.~Antonenko,  W.~Scheid, and H.Q.~Zhang,
%Eur. Phys. J. A {\bf 47}, 38 (2011).
\bibitem{EPJSub1}        V.V.~Sargsyan, G.G.~Adamian, N.V.~Antonenko,
W.~Scheid, and H.Q.~Zhang, Phys. Rev. C {\bf 84}, 064614 (2011);
 {\it ibid}  {\bf 85}, 024616 (2012);  {\it ibid}  {\bf 86}, 014602 (2012);
 {\it ibid}  {\bf 86}, 034614 (2012).
%\bibitem{Conf}      V.V.~Sargsyan~{\it et al.},
%%, G.G.~Adamian, N.V.~Antonenko,  W.~Scheid, and H.Q.~Zhang,
% J. of Phys.: Conf. Ser. {\bf 282}, 012001 (2011);  EPJ Web Conf. {\bf 17}, 04003 (2011).
\bibitem{Ayik333} S.~Ayik, B.~Yilmaz, and D.~Lacroix,  Phys. Rev. C {\bf 81}, 034605 (2010).
\bibitem{PRCPOP}       A.S.~Zubov, V.V.~Sargsyan,  G.G.~Adamian, and N.V.~Antonenko,
                        Phys. Rev. C {\bf 84}, 044320 (2011);
 A.S.~Zubov, V.V.~Sargsyan,  G.G.~Adamian,   N.V.~Antonenko, and W.~Scheid,
                        Phys. Rev. C {\bf 81}, 024607 (2010); Phys. Rev. C {\bf 82}, 034610 (2010);
R.A.~Kuzyakin, V.V.~Sargsyan, G.G.~Adamian, N.V.~Antonenko, E.E.~Saperstein,
and S.V.~Tolokonnikov, Phys. Rev. C {\bf 85}, 034612 (2012).
\bibitem{Hofman}        H.~Hofmann, Phys. Rep.  {\bf 284}, 137 (1997);
%;                       C.~Rummel and H.~Hofmann, Nucl. Phys. A {\bf 727}, 24 (2003).
%\bibitem{Ayik}
%N.~Takigawa, S.~Ayik, K.~Washiyama, and S.~Kimura, Phys. Rev. C {\bf 69}, 054605 (2004);
S.~Ayik, B.~Yilmaz, A.~Gokalp, O.~Yilmaz, and N.~Takigawa, Phys. Rev. C {\bf 71}, 054611 (2005);
%
%\bibitem{our}          V.V.~Sargsyan, Z.~Kanokov, G.G.~Adamian, N.V.~Antonenko, and W.~Scheid,
%                        Phys. Rev. C {\bf 80}, 034606 (2009); Phys. Rev. C {\bf 80}, 047603 (2009);
                        V.V.~Sargsyan, Z.~Kanokov, G.G.~Adamian, and N.V.~Antonenko,
                        Part. Nucl. {\bf 41}, 175 (2010);
%
%\bibitem{Hupin}
G.~Hupin and D.~Lacroix,  Phys. Rev. C {\bf 81}, 014609 (2010).
\bibitem{SSzilner} S.~Szilner {\it et al.},
Phys. Rev. C {\bf 76}, 024604 (2007);
S.~Szilner {\it et al.}, Phys. Rev. C {\bf 84}, 014325 (2011);
L.~Corradi {\it et al.}, Phys. Rev. C {\bf 84}, 034603 (2011).
%L.~Corradi, G.~Pollarolo, and S.~Szilner, J. Phys. G  {\bf 36}, 113101  (2009);
%S.~Szilner {\it et al.}, J. of Phys.: Conf. Ser. {\bf 282}, 012021 (2011).
\bibitem{obzor}     G.G.~Adamian, A.K.~Nasirov, N.V.~Antonenko, and R.V.~Jolos,
Phys. Part. Nucl. {\bf 25}, 583 (1994);
K.~Washiyama, D.~Lacroix, and S.~Ayik,  Phys. Rev. C {\bf 79}, 024609 (2009);
S.~Ayik, K.~Washiyama, and D.~Lacroix,  Phys. Rev. C {\bf 79}, 054606 (2009).

\bibitem{Ram}  S.~Raman, C.W.~Nestor, Jr, and P.~Tikkanen, At. Data Nucl. Data Tables {\bf 78}, 1 (2001).
\bibitem{Kanada} Y.~Kanada-En'yo, H.~Horiuchi, and A.~Ono,
Phys. Rev. C {\bf 52}, 628 (1995); A~Dote, H.~Horiuchi, and Y.~Kanada-En'yo,
Phys. Rev. C {\bf 56}, 1844 (1997).


%%%%%              Experiment          %%%%%%%%%%%%%%%%

\bibitem{B11Bi209}       L.R.~Gasques, D.J.~Hinde, M.~Dasgupta, A.~Mukherjee, and R.G.~Thomas,
                         Phys. Rev. C {\bf 79}, 034605 (2009).
\bibitem{B11Tb159}       A.~Mukherjee {\it et al.}, Phys. Lett. B {\bf 636}, 91 (2006).
%\bibitem{Ram}            S.~Raman, C.W.~Nestor, Jr, and P.~Tikkanen, At. Data Nucl. Data Tables {\bf 78}, 1 (2001).
\bibitem{Be9Bi209}       C.~Signorini {\it et al.}, Nucl. Phys. {\bf A735}, 329 (2004); M.~Dasgupta,
D.J.~Hinde, S.L.~Sheehy, and B.~Bouriquet,  Phys. Rev. C {\bf 81}, 024608 (2010).
\bibitem{Be9Pb208}       M.~Dasgupta  {\it et al.}, Phys. Rev. Lett. {\bf 82}, 1395 (1999).
\bibitem{Be9Sm144}       P.R.S.~Gomes {\it et al.}, Phys. Rev. C {\bf 73}, 064606 (2006).
\bibitem{Be9Sn124}       V.V.~Parkar {\it et al.}, Phys. Rev. C {\bf 82}, 054601 (2010).
\bibitem{Be9Y89}         C.S.~Palshetkar {\it et al.}, Phys. Rev. C {\bf 82}, 044608 (2010).
\bibitem{Be9Zn64}        P.R.S.~Gomes {\it et al.}, Phys. Rev. C {\bf 71}, 034608  (2005).
\bibitem{Be9Al27}        G.V.~Marti {\it et al.}, Phys. Rev. C {\bf 71}, 027602  (2005).
\bibitem{Li7Al27}        I.~Padron {\it et al.}, Phys. Rev. C {\bf 66}, 044608  (2002).
%\bibitem{Be9U238}        V.~Fekou-Youmbi {\it et al.}, J. Phys. G: Nucl. Part. Phys. {\bf 23} 1259-1266 (1997).
\bibitem{Li7Au197}       Sh.~Thakur {\it et al.},  EPJ Web  Conf.   {\bf 17}, 16017 (2011).
\bibitem{Li6Bi209}       M.~Dasgupta {\it et al.}, Phys. Rev. C {\bf 70}, 024606 (2004).
\bibitem{Li6Pb208}       Y.W.~Wu {\it et al.}, Phys. Rev. C {\bf 68}, 044605 (2003).
\bibitem{Li6Pt198}       A.~Shrivastava {\it et al.}, Phys. Rev. Lett. {\bf 103}, 232702 (2009).
\bibitem{Li6Sm144}       P.K.~Rath {\it et al.}, Phys. Rev. C {\bf 79}, 051601(R) (2009).
\bibitem{Li6Zn64}        D.~Torresi {\it et al.},  EPJ Web  Conf.  {\bf 17}, 16018 (2011).
\bibitem{Li7Ho165}       V.~Tripathi {\it et al.}, Phys. Rev. Lett. {\bf 88}, 172701 (2002).
\bibitem{He4Bi211}       J.J.~Kolata {\it et al.}, Eur. Phys. J. A {\bf 13}, 117-121 (2002).
\bibitem{He4Zn64}        M.~Fisichella {\it et al.},  EPJ Web  Conf.  {\bf 17}, 16003 (2011).
\bibitem{Li9Zn70}        W.~Loveland {\it et al.}, Phys. Rev. C {\bf 74}, 064609 (2006);
                         W.~Loveland,  EPJ Web Conf. {\bf 17}, 02003 (2011).
\bibitem{He4Au197}       A.~Lemasson {\it et al.}, Phys. Rev. Lett. {\bf 103}, 232701 (2009);
                         A.~Lemasson  EPJ Web  Conf.   {\bf 17}, 01003 (2011).
\bibitem{Li9Pb208}       A.M.~Vinodkumar {\it et al.}, Phys. Rev. C {\bf 80}, 054609 (2009).
\bibitem{Maximka}        A.~Diaz-Torres, J. Phys. G {\bf 37}, 075109 (2010).
\bibitem{PLATYPUS}        A.~Diaz-Torres, Comp. Phys. Comm. {\bf 182}, 1100 (2011).
\bibitem{Luong}          D.H.~Luong {\it et al.}, Phys. Lett. B {\bf 695}, 105 (2011);
                                                  EPJ Web Conf. {\bf 17}, 03002 (2011).
%                        M.~Dasgupta {\it et al.}, Nucl. Phys.  {\bf A834}, 147c (2010).
\bibitem{Gomesnew}        P.R.S.~Gomes, private communication  (2012).
\bibitem{Timmers}    H.~Timmers, J.R.~Leigh, M.~Dasgupta, D.J.~Hinde,
R.C.~Lemmon, J.C.~Mein, C.R.~Morton, J.O.~Newton, and N.~Rowley,
 Nucl. Phys.  {\bf A584}, 190 (1995).
\bibitem{Zhang}    H.Q.~Zhang, F.~Yang, C.~Lin, Z.~Liu, and Y.~Hu, Phys. Rev. C {\bf 57}, R1047 (1998);
Inter. workshop on {\it Nuclear reactions and beyond, Wold Scientific (2000), p.95};
F.~Yang, C.J.~Lin, X.K.~Wu, H.Q.~Zhang,
 C.L.~Zhang, P.Zhou, and Z.H.~Liu, Phys. Rev. C {\bf 77}, 014601 (2008).
\bibitem{Sonzogni}    A.A.~Sonzogni, J.D.~Bierman, M.P.Kelly, J.P.Lestone, J.F.Liang, and
R.~Vandenbosch, Phys. Rev. C {\bf 57}, 722 (1998).
\bibitem{Sinha}    S.~Sinha, M.R.~Pahlavani, R.~Varma, R.K.~Choudhury, B.K.~Nayak, and A.~Saxena,
Phys. Rev. C {\bf 64}, 024607 (2001).
\bibitem{Piasecki}    E.~Piasecki {\it et al.}, Phys. Rev. C {\bf 65}, 054611 (2002);
E.~Piasecki {\it et al.}, Phys. Rev. C {\bf 80}, 054613 (2009);
E.~Piasecki {\it et al.}, Phys. Rev. C {\bf 85}, 054604 (2012); {\it ibid} {\bf 85}, 054608 (2012).


\end{thebibliography}
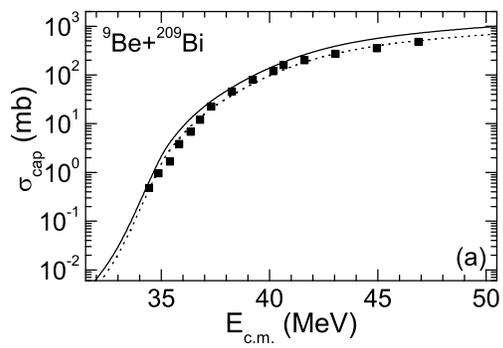
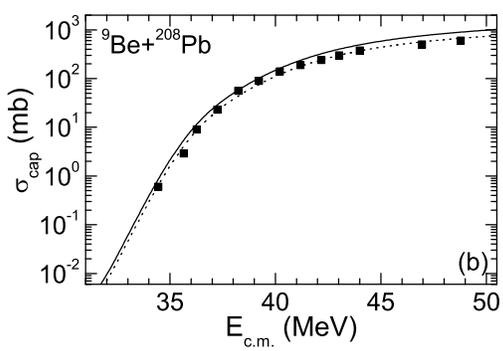
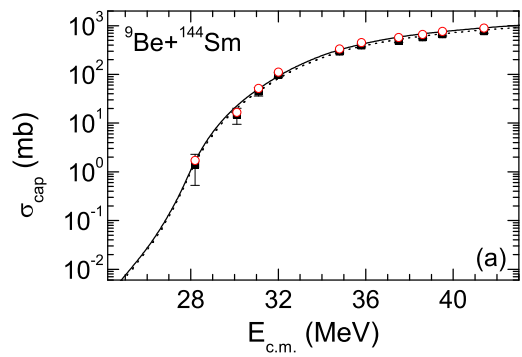
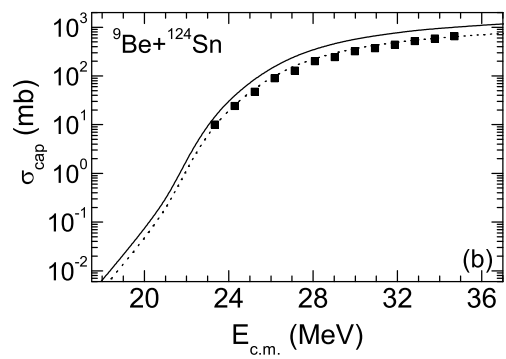
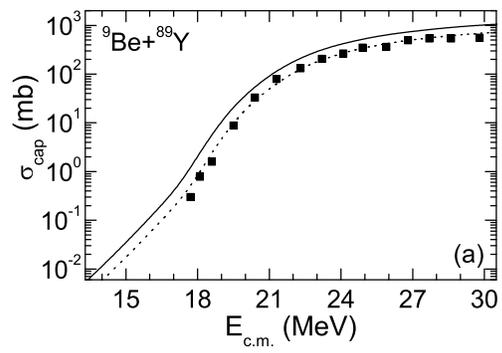
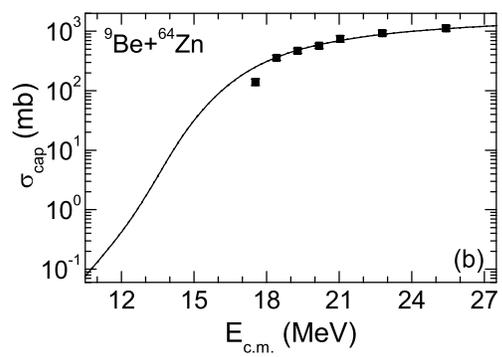
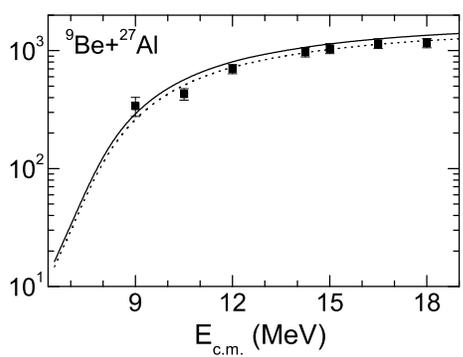
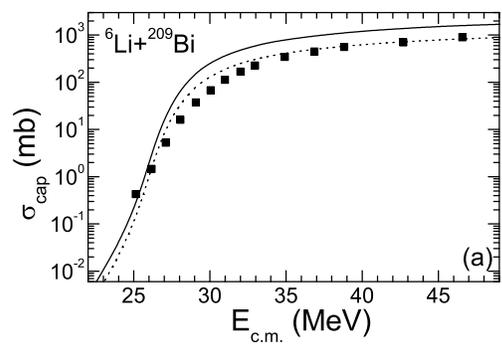
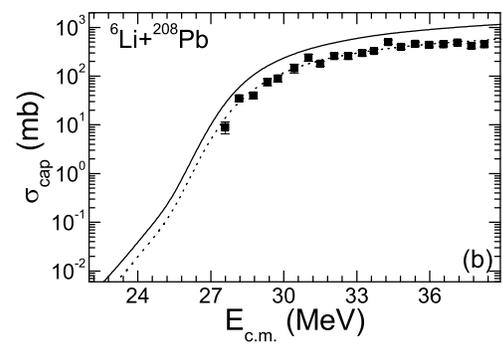
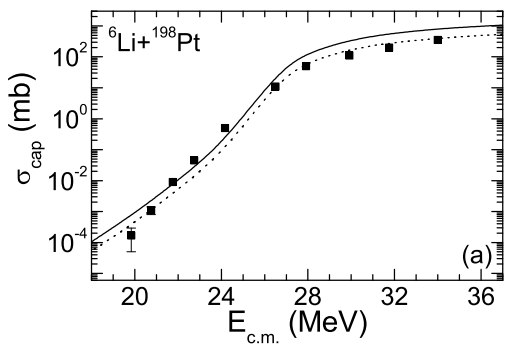
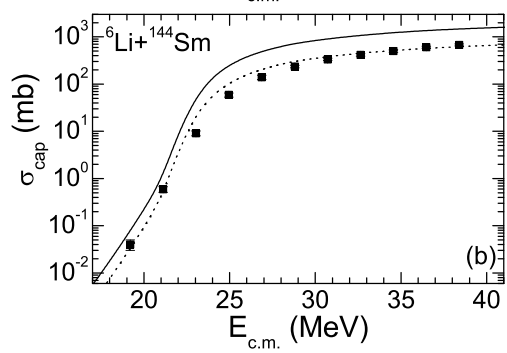
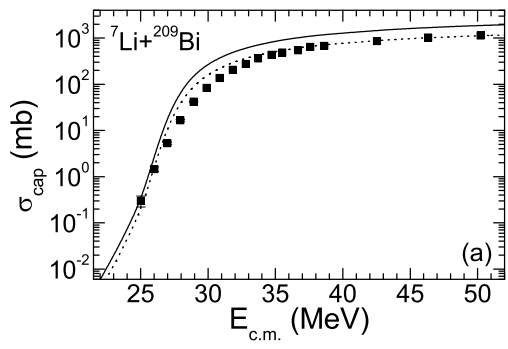
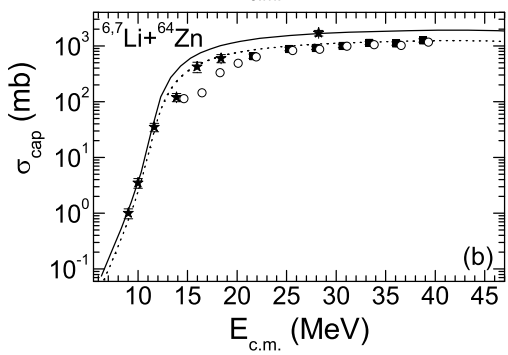
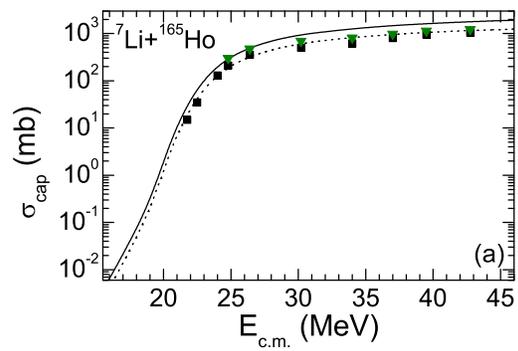
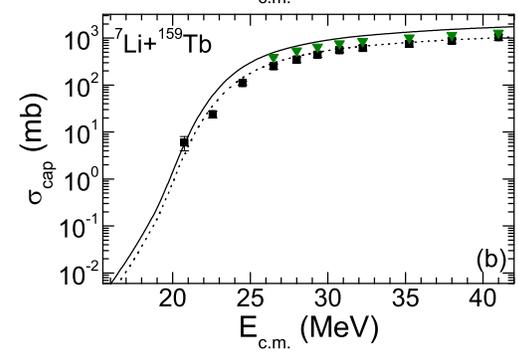
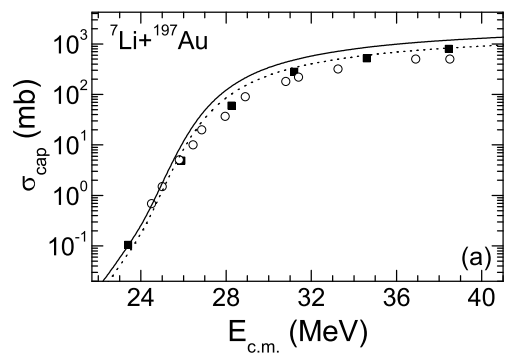
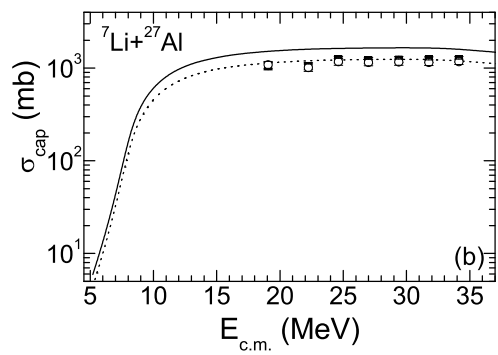
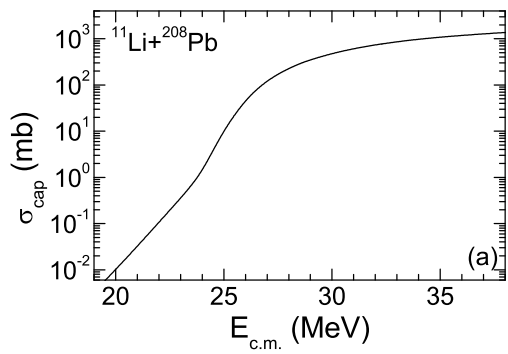
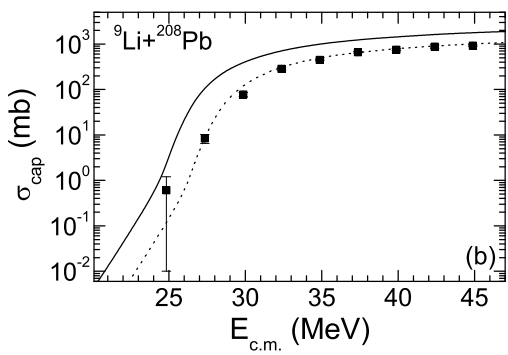
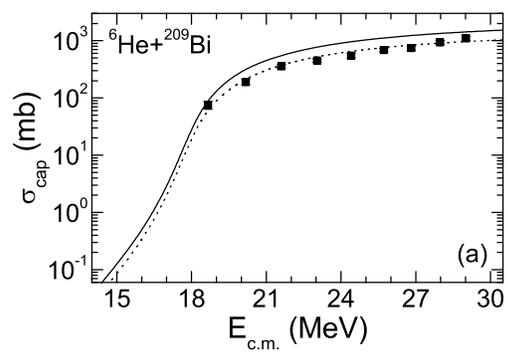
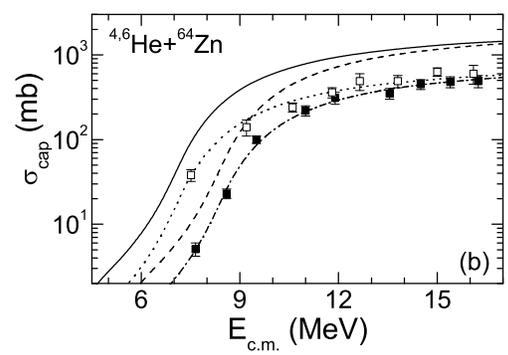
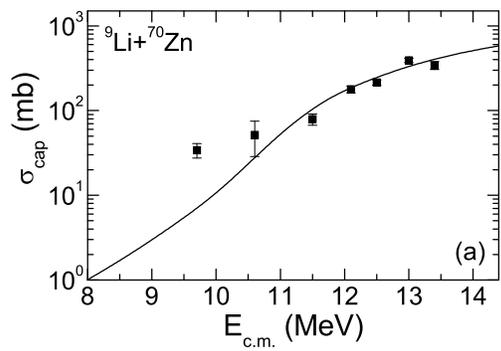
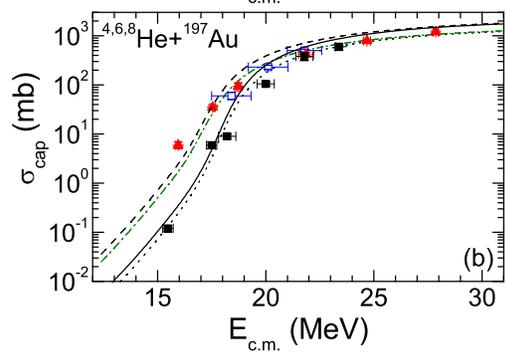
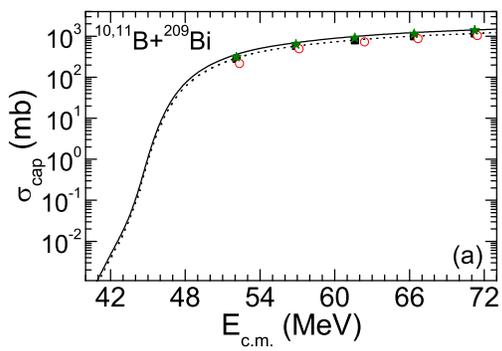
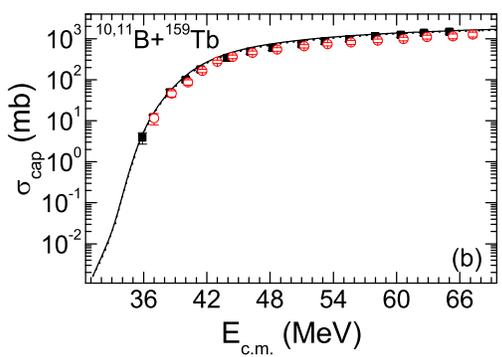
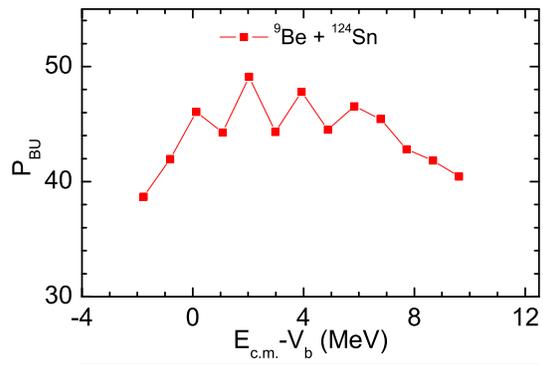
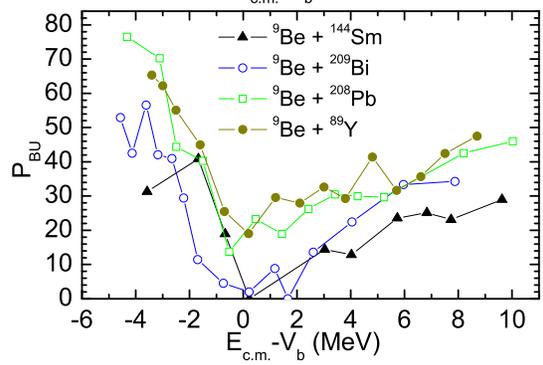
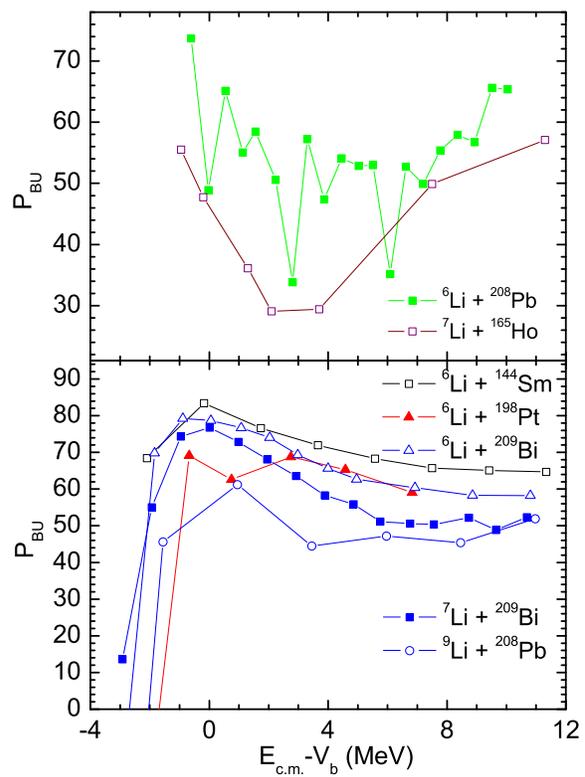
\end{document}